\documentclass[traditabstract]{aa}

\usepackage{color}
\usepackage{graphicx}
\usepackage{amssymb}
\usepackage{amsmath}
\usepackage{bbold}
\usepackage{url}
\usepackage{natbib}
\bibpunct{(}{)}{;}{a}{}{,}

\newcommand{\onelargefig}[3]{%
  \begin{figure*}%
    \centerline{\resizebox{\hsize}{!}{\includegraphics*{#1}}}%
    \caption{#3}\label{#2}%
  \end{figure*}%
}
\newcommand{\twofig}[4]{%
  \begin{figure*}%
    \centerline{\resizebox{\hsize}{!}{\includegraphics*{#1} \,%
        \includegraphics*{#2}}}%
    \caption{#4}\label{#3}%
  \end{figure*}%
}
\newcommand{\threefig}[5]{%
  \begin{figure*}%
    \centerline{\resizebox{\hsize}{!}{\includegraphics*{#1} \,%
        \includegraphics*{#2} \,%
        \includegraphics*{#3}}}%
    \caption{#5}\label{#4}%
  \end{figure*}%
}

\newcommand{\sect}[1]{Sect.~\ref{#1}}

\newcommand{\fig}[1]{Fig.~\ref{#1}}

\newcommand{\eq}[1]{Eq.~(\ref{#1})}

\newcommand{\tab}[1]{Table~\ref{#1}}

\newcommand{\lmax}{\ifmmode \ell_{\mathrm{max}}\else $\ell_{\mathrm{max}}$\fi}
\newcommand{\npix}{\ifmmode n_{\mathrm{pix}}\else $n_{\mathrm{pix}}$\fi}
\newcommand{\nside}{\ifmmode n_{\mathrm{side}}\else $n_{\mathrm{side}}$\fi}
\newcommand{\lm}{\ifmmode \ell m\else $\ell m$\fi}
\newcommand{\alm}{\ifmmode a_{\ell m}\else $a_{\ell m}$\fi}
\newcommand{\GHz}{\ifmmode \,\mathrm{GHz}\else GHz\fi}
\newcommand{\order}[1]{${{\cal O}\! \left( #1 \right)}$}

\newcommand{\planck}{\emph{Planck}}
\newcommand{\keck}{\emph{Keck}}
\newcommand{\cc}{compressed convolution}
\newcommand{\beq}{\begin{equation}}
\newcommand{\eeq}{\end{equation}}
\newenvironment{referee}{\bf}{}
\newcommand{\bref}{\begin{referee}}
\newcommand{\eref}{\end{referee}}

\begin{document}

\title{Compressed convolution}

\titlerunning{Compressed convolution}

\author{Franz Elsner\inst{1,2}
  \and
  Benjamin D. Wandelt\inst{2,3}}

\institute{Department of Physics and Astronomy, University College
  London, London WC1E 6BT, U.K.\\
  \email{f.elsner@ucl.ac.uk}
  \and
  Institut d'Astrophysique de Paris, UMR 7095, CNRS - Universit\'e
  Pierre et Marie Curie (Univ Paris 06), 98 bis blvd Arago, 75014
  Paris, France
  \and
  Departments of Physics and Astronomy, University of Illinois at
  Urbana-Champaign, Urbana, IL 61801, USA}

\date{Received \dots / Accepted \dots}

\abstract{We introduce the concept of \cc, a technique to convolve a
  given data set with a large number of non-orthogonal kernels. In
  typical applications our technique drastically reduces the effective
  number of computations. The new method is applicable to convolutions
  with symmetric and asymmetric kernels and can be easily controlled
  for an optimal trade-off between speed and accuracy. It is based on
  linear compression of the collection of kernels into a small number
  of coefficients in an optimal eigenbasis. The final result can then
  be decompressed in constant time for each desired convolved output.
  The method is fully general and suitable for a wide variety of
  problems. We give explicit examples in the context of simulation
  challenges for upcoming multi-kilo-detector cosmic microwave
  background (CMB) missions. For a CMB experiment with \order{10\,000}
  detectors with similar beam properties, we demonstrate that the
  algorithm can decrease the costs of beam convolution by two to three
  orders of magnitude with negligible loss of accuracy. Likewise, it
  has the potential to allow the reduction of disk space required to
  store signal simulations by a similar amount.  Applications in other
  areas of astrophysics and beyond are optimal searches for a large
  number of templates in noisy data, e.g.\ from a parametrized family
  of gravitational wave templates; or calculating convolutions with
  highly overcomplete wavelet dictionaries, e.g.\ in methods designed
  to uncover sparse signal representations.}


\maketitle

\section{Introduction}
\label{sec:intro}

Convolution is a very common operation in processing pipelines of
scientific data sets. For example, in the analysis of cosmic microwave
background (CMB) radiation experiments, convolutions are used to
improve the detection of point sources
\citep[e.g.,][]{1998ApJ...500L..83T, 2000MNRAS.315..757C}, in the
search for non-Gaussian signals on the basis of wavelets
\citep[e.g.,][]{2001MNRAS.327..813B, 2002MNRAS.336...22M}, during
mapmaking \citep[e.g.,][]{1997ApJ...480L..87T, 2001A&A...372..346N},
or Wiener filtering \citep{2013A&A...549A.111E}.

Convolution for data simulation presents similar if not greater
challenges: the current and next generations of CMB experiments are
nearly photon-noise limited. The only way to reach the sensitivity
required to detect and resolve B-modes or to resolve the
Sunyaev-Zel'dovich effect of clusters of galaxies over large fractions
of sky is to build detector arrays with $N\sim10^{2}-10^{4}$
detectors. Simulating the signal for these experiments requires
convolving the same input sky with $N$ different and often quite
similar kernels.

In the simplest case, when the convolution kernel is azimuthally
symmetric, convolution involves the computation of spherical harmonic
transformations. Although highly optimized implementations exist
(e.g., libsharp, \citealt{2013A&A...554A.112R}, the default back end
in the popular HEALPix library, \citealt{2005ApJ...622..759G}),
spherical harmonic transformations are numerically expensive and may
easily become the bottleneck in data simulation and processing
pipelines.

Even more critical is the more realistic setting when the kernels are
anisotropic (e.g., when modeling the physical optics of a CMB
experiment or when performing edge or ridge detection with curvelets
or steerable filters, e.g., \citealt{2006ApJ...652..820W,
  2007ITSP...55..520M}) In this case, the cost of convolution
additionally scales with the degree of azimuthal structure in the
kernel \citep{2001PhRvD..63l3002W} and the convolution output is
parametrized in terms of three Euler angles each taking $\sim L$
distinct values, where $L$ is the bandlimit of the convolution
output. Storing thousands of such objects, one for each beam, requires
storage capacity approaching the peta-byte scale.

In this paper, we show that regardless of the details of the
convolution problem, or the algorithm used for performing the
convolution, the computational costs and storage requirements
associated with multiple convolutions can be considerably reduced as
long as the set of convolution kernels contains linearly compressible
redundancy. Our approach exploits the linearity of the convolution
operation to represent the set of convolution kernels in terms of an
often much smaller set of optimal basis kernels. We demonstrate that
this approach can greatly accelerate several examples taken from CMB
data simulation and analysis.

Approaches based on singular value decompositions (SVD) have already
proven very successful in observational astronomy to correct imaging
data for spatially varying point spread functions
\citep[e.g.,][]{2001ASPC..238..269L, 2002SPIE.4847..167L}. Likewise,
SVDs have been used to accelerate the search for gravitational wave
signatures \citep[e.g.,][]{2010PhRvD..82d4025C} using precomputed
templates \citep{lrr-2005-3}. In this paper, we show these
methods to be special cases of a more general approach that returns a
signal-to-noise eigenbasis that achieves optimal acceleration and
compression for a given accuracy goal.

The paper is organized as follows. In \sect{sec:method}, we introduce
the mathematical foundations of our method. Using existing spaceborne
and ground based CMB experiments as an example, we then analyze the
performance of the \cc\ scheme when applied to the beam convolution
problem (\sect{sec:example}). After outlining the scope of our
algorithm in \sect{sec:scope}, we summarize our findings in
\sect{sec:summary}.

\section{Method}
\label{sec:method}

Starting from the defining equation of the convolution integral, we
first review the basic concept of the algorithm. Given a raw data map
$d(\vec{x})$, the convolved object (time stream, map) $s(\vec{x})$ is
derived by convolution with a kernel $K(\vec{x}, \vec{y})$,
\beq
\label{eq:conv}
s(\vec{x}) = \int K(\vec{x}, \vec{y}) d(\vec{y}) d\vec{y} \, .
\eeq
Note that, without loss of generality, we focus on the convolution of
two-dimensional data sets in this paper.

\subsection{Overview}

In practice, a continuous signal is usually measured only on a finite
number of discrete pixels. We therefore approximate the integral in
\eq{eq:conv} by a sum in what follows,
\begin{align}
  \label{eq:conv_discrete}
  s_i &= \sum_j K_{i, j} d_j \nonumber \\
  &=  \left(R_{i} k\right) ^{\dagger}d  \, .
\end{align}
For our subsequent analysis, we introduced the operator $R$ in the
latter equation such that $R_{i} k$ is the $i^{\mathrm{th}}$ row of
the convolution matrix, constructed from the convolution kernel $K$.

For any complete set of basis functions $\{\phi^1, \dots, \phi^N\}$,
there exists a unique set of coefficients $\{\lambda^1, \dots,
\lambda^N\}$, such that
\beq
K_{i, j} = \sum_k \lambda^k \widehat{K}(\phi^k)_{i, j} \, ,
\eeq
i.e., we do a basis transformation of the kernel from the standard
basis to the basis given by the $\{\phi^k\}$.

Taking advantage of the
linearity of the convolution operation, \eq{eq:conv_discrete} can then
be transformed to read
\begin{align}
  \label{eq:conv_decomposition}
  s_i &= \sum_j \left( \sum_k \lambda^k \widehat{K}(\phi^k)_{i, j}
  \right) d_j \nonumber \\
      &= \sum_k \lambda^k \left( \sum_j \widehat{K}(\phi^k)_{i, j} \,
  d_j \right) \nonumber \\
      &= \sum_k \lambda^k s_i^k \, ,
\end{align}
where the $s^k$ are the raw input map convolved with the
$k^{\mathrm{th}}$ mode of the basis functions themselves. That is, the
final convolution outputs are now expressed in terms of a weighted sum
of individually convolved input maps with a set of basis kernels.

We note that for a single convolution operation, the decomposition of
the convolution kernel into multiple basis functions in
\eq{eq:conv_decomposition} cannot decrease the numerical costs of the
operation. However, potential performance improvements can be realized
if multiple convolutions are to be calculated, as we will discuss in
the following.

Consider the particular problem where a single raw map $d$ should be
convolved with $n_{\mathrm{tot}}$ different convolution kernels, i.e.,
we want to compute
\beq
s_i^{(n)} = \sum_j K_{i, j}^{(n)} d_j \, ,
\eeq
where we introduced the kernel ID $n \in \{1, \dots,
n_{\mathrm{tot}}\}$ as a running index.

Applying the kernel decomposition into a common set of basis
functions, \eq{eq:conv_decomposition} now reads
\beq
\label{eq:kernel_expansion}
s_i^{(n)} = \sum_k \lambda^{(n), k} s_i^k \, .
\eeq
This finding builds the foundation of our fast algorithm: the
numerically expensive convolution operations are applied only to a
limited number of basis modes used in the expansion. The computational
cost is therefore largely independent of the total number of kernels,
$n_{\mathrm{tot}}$, since each individual solution is constructed very
efficiently via a simple linear combination out of a set of
precomputed convolution outputs.

\subsection{Optimal kernel expansion}

For the kernel decomposition in \eq{eq:kernel_expansion} to be useful
in practice, we have to restrict the total number of basis modes for
which the convolution is calculated explicitly. To find the optimal
expansion, i.e., the basis set with the smallest number of modes for a
predefined truncation error, we first define the weighted sum of the
expected covariance of all the elements of the convolution output
\beq
\label{eq:error_measure}
\sigma^2 = \left \langle \sum_{(n)} \sum_{i, \, i^{\prime}}
s_i^{(n)} N^{(n)\,-1}_{i \, i^{\prime}} s_{i^{\prime}}^{(n)} \right \rangle \, .
\eeq

Here, we have introduced a real symmetric weighting matrix, $N^{(n)}$,
which allows us to specify what aspects of the convolved maps we
require to be accurate. For the case of convolving to simulate CMB
data, a natural choice for $N^{(n)}$ would be the noise covariance for
the $n^{\mathrm{th}}$ channel. It ensures that any given channel will
be simulated at sufficient accuracy and that after the addition of
instrumental noise, the statistics of the resulting simulation are
indistinguishable from an exact simulation.

It is now easy to see how to decompose the kernels into a basis such a
way as to concentrate the largest amount of variance in the first
basis elements. Define the Hermitian matrix
\begin{align}
\label{eq:weighted_covariance}
M_{nm} &= \left \langle \sum_{i, \, i^{\prime}, \, i^{\prime \prime}}
N^{(n) \, -\frac{1}{2}}_{i^\prime \, i} s_i^{(n)} N^{(m) \,
  -\frac{1}{2}}_{i^{\prime} \, i^{\prime \prime}}
s_{i^{\prime \prime}}^{(m)} \right \rangle \nonumber \\
& = \sum_{i, \, i^{\prime}} \left[ \left( N^{(n) \,
    -\frac{1}{2}} \right)^{\dagger} N^{(m) \, -\frac{1}{2}} \right]_{i
  \, i^{\prime}} \left( R_{i^{\prime}}
k^{(m)} \right) C \left( R_{i} k^{(n)} \right)^{\dagger} \, ,
\end{align}
where $C$ is the covariance of the input signal and $N^{(n)\,\frac{1}{2}}$
is any matrix such that $(N^{(n) \, \frac{1}{2}})^{\dagger}  N^{(n) \,
  \frac{1}{2}} = N^{(n)}$.

Then we can rewrite the scalar \eq{eq:error_measure} as a matrix trace
over the kernel IDs
\begin{align}
  \label{eq:error_criterion}
  \sigma^2 &= \left\langle \mathrm{tr} \left( M \right) \right\rangle
  \nonumber \\
  &= \sum_{n, \, i, \, i^{\prime}} N^{(n)\,-1}_{i \, i^{\prime}}
  (R_{i^{\prime}} k^{(n)}) C (R_{i} k^{(n)})^{\dagger} \,.
\end{align}

Since the matrix in \eq{eq:error_criterion} is Hermitian, its ordered
diagonal elements cannot decrease faster than its ordered eigenvalues
by Schur's theorem. Finding the eigensystem of $M$ therefore results
in the kernel decomposition that converges faster than any other
decomposition to the result of the direct computation. In other words,
the decomposition is optimal because discarding the eigenmodes with
the smallest eigenvalues results in the smallest possible change in
the overall signal power.

If we denote the eigenvectors of $M$ by $u^{r}$, with corresponding
eigenvalues $\nu_{r}$, the optimal compression kernel eigenmodes are
given by $\phi^{(n)}_{i} = \sum_{m} u^{(n)}_{(m)} k_{i}^{(m)}$, and
the mean square truncation error is the sum of the truncated
eigenvalues.

Considering the CMB case of a convolution on the sphere with
azimuthally symmetric convolution kernels and multipole-dependent
diagonal weights, $N_{\ell}$, \eq{eq:weighted_covariance} simplifies
to
\beq
M_{nm} = \sum_{\ell} \frac{2\ell+1}{4\pi} \left(
\frac{\mathcal{C}_{\ell}}{\sqrt{N^{(n)}_{\ell} N^{(m)}_{\ell}}}
\right) K^{(n)}_{\ell} K^{(m)}_{\ell} \, ,
\eeq
and \eq{eq:error_criterion} becomes
\beq
\sigma^{2} = \sum_{\ell, \, n} \frac{2\ell+1}{4\pi} \left(
\frac{\mathcal{C}_{\ell}}{N^{(n)}_{\ell}} \right) K^{(n)}_{\ell}
K^{(n)}_{\ell}\, ,
\eeq
which clearly shows the signal-to-noise weighting at work.

Note, that the expression for the variance can be promoted to a matrix
in a dual way,
\beq
\label{eq:dual_{covmatrix}}
M_{ll^{\prime}} = \sum_{n}\sqrt{ \frac{2 \ell + 1}{4 \pi}
\frac{\mathcal{C}_{\ell}}{ N^{(n)}_{\ell}}}\sqrt{\frac{2 \ell^{\prime} + 1}{4 \pi}
\frac{\mathcal{C}_{\ell^{\prime}}}{ N^{(n)}_{\ell^{\prime}}}} K_{\ell}^{(n)}
K_{\ell^{\prime}}^{(n)} \, ,
\eeq
which gives rise to an alternative way to calculate the optimal
compression basis.

This dual approach will be computationally more convenient than the
other approach if the number of kernels is larger than the number of
multipoles in the $\ell$-range considered. The resulting compression
scheme will be identical in both cases. This is so because both
approaches are optimal by Schur's theorem and each gives a unique
answer if none of the eigenvalues are degenerate\footnote{If some
  eigenvalues do happen to be degenerate then the solutions will
  differ in ways that are not relevant to the compression
  efficiency.}.

\subsection{Truncation error estimates}

In case the kernels are of similar shape, or differ only in regimes
that are irrelevant due to low signal-to-noise, the eigenvalues of the
individual modes will decrease quickly. As a result, we can truncate
the expansion in \eq{eq:kernel_expansion} at $n_{\mathrm{modes}} \ll
n_{\mathrm{tot}}$. This will induce a mean square truncation in the
weighted variance of the convolution products of
$\sum_{r=n_{\mathrm{modes}}+1}^{n_{\mathrm{tot}}} \nu_{r}$.

The error $\Delta K$ introduced by the truncation can be calculated
for each kernel explicitly,
\beq
\label{eq:error}
\Delta K_{i, j} = \sum_{k = n_{\mathrm{modes}} + 1}^{n_{\mathrm{tot}}}
\lambda^k \widehat{K}(\phi^k)_{i, j} \, .
\eeq

For the convolution of a data set with power spectrum
$\mathcal{C}_{\ell}$ on the sphere, for example, the mean square error
will then amount to
\beq
\label{eq:total_error}
\sigma^2_{\mathrm{total}} = \sum_{\ell = 0}^{\lmax} \frac{(2\ell +
  1)}{4\pi} \, \Delta K_{\ell}^2 \, \mathcal{C}_{\ell} \, ,
\eeq
where $\Delta K_{\ell}$ is the expansion of the beam truncation error
into Legendre polynomials.

\subsection{Connection to the SVD}

While \eq{eq:error_criterion} provides us with the optimal kernel
decomposition, the power spectrum of the data or their noise
properties to construct the kernel weights may not necessarily be
known in advance. For uniform weightings, $N \propto \mathbb{1}$, and
assuming a flat signal power spectrum, the equation simplifies and we
obtain the mode expansion from a singular value decomposition of the
collection of kernels.

Although not strictly optimal, we note that it is possible to obtain
good results with this simplified approach in practice. To compute the
kernel expansion, we reshape the convolution kernels into
one-dimensional arrays of length $m$ and arrange them into a common
matrix $T$, with size $n_{\mathrm{tot}} \times m$. The singular value
decomposition of this matrix,
\beq
\label{eq:svd}
T = U D V^{\dagger} \, ,
\eeq
computes the $n_{\mathrm{tot}} \times n_{\mathrm{tot}}$ matrix U, the
$n_{\mathrm{tot}} \times m$ matrix D, and the $m \times m$ matrix V.
The decomposition then provides us with a set of basis functions,
returned in the columns of $V$. Their relative importance is indicated
by the entries of the diagonal matrix $D$, and their individual
coefficients $\lambda$ are stored in $U$.

\subsection{Summary}

In summary, the individual steps of the algorithm are as follows: We
first find the eigenmode decomposition of the set of convolution
kernels using either the optimal expansion criterion or a simplified
singular value decomposition. Then, we identify the number of modes to
retain to comply with the accuracy goal. As a next step, we perform
the convolution of the input map for each eigenmode separately. To
obtain the final results, we compute the linear combination of the
convolved maps with optimal weights for each kernel.

It is worth noting that \cc\ can never increase the computational time
required for convolution, except possibly for some overhead of
sub-leading order, attributed to the calculation of the optimal kernel
expansion (this computation has to be done only once for a given set
of kernels). This can be seen explicitly in the worst case scenario of
strictly orthogonal kernels: all modes must be retained and the method
becomes equivalent to the brute force approach.

\section{Application to CMB experiments}
\label{sec:example}

After having outlined the basic principle of the algorithm, we now
analyze the performance of the method when applied to the beam
convolution operation of current CMB experiments.

\subsection{Planck}

We use the third generation CMB satellite experiment
\planck\ \citep{2011A&A...536A...1P} as a first example to illustrate
the application of the algorithm. We make use of the $217 \GHz$ HFI
instrument \citep{2011A&A...536A...4P} and consider the beam
convolution problem of CMB simulations. Azimuthally symmetrized beam
functions for the six individual detectors at that frequency are
available from the reduced instrument model
\citep{2013arXiv1303.5068P}.

A comparison of the eigenvalues of a singular value decomposition
reveals that the beam shapes are sufficiently similar to be
represented with only a limited number of basis functions
(\fig{fig:pl_beams}). As shown in \fig{fig:pl_modes1}, selecting the
first three eigenmodes for a reconstruction is sufficient to represent
the beams to an accuracy of the order \order{10^{-3}}, better than the
typical precision to which the beams are known.

To illustrate the impact of the weighting scheme, we also show the
resulting eigenmodes using the optimal kernel expansion
(\eq{eq:dual_{covmatrix}}) in \fig{fig:pl_modes2}. Here, we assumed a
white noise power spectrum, $N_{\ell} = const.$, in combination with a
signal covariance of $\mathcal{C}_{\ell} \propto 1/(\ell \, (\ell +
1))$, reflecting the approximate scaling behavior of the CMB power
spectrum.

We chose the beam with the largest reconstruction error for an
explicit test on simulated CMB signal maps. In \fig{fig:pl_error}, we
plot the difference map computed from the brute force beam convolution
and the \cc\ with three eigenmodes. A power spectrum analysis confirms
that the truncation induced errors are clearly subdominant on all
angular scales.

\twofig{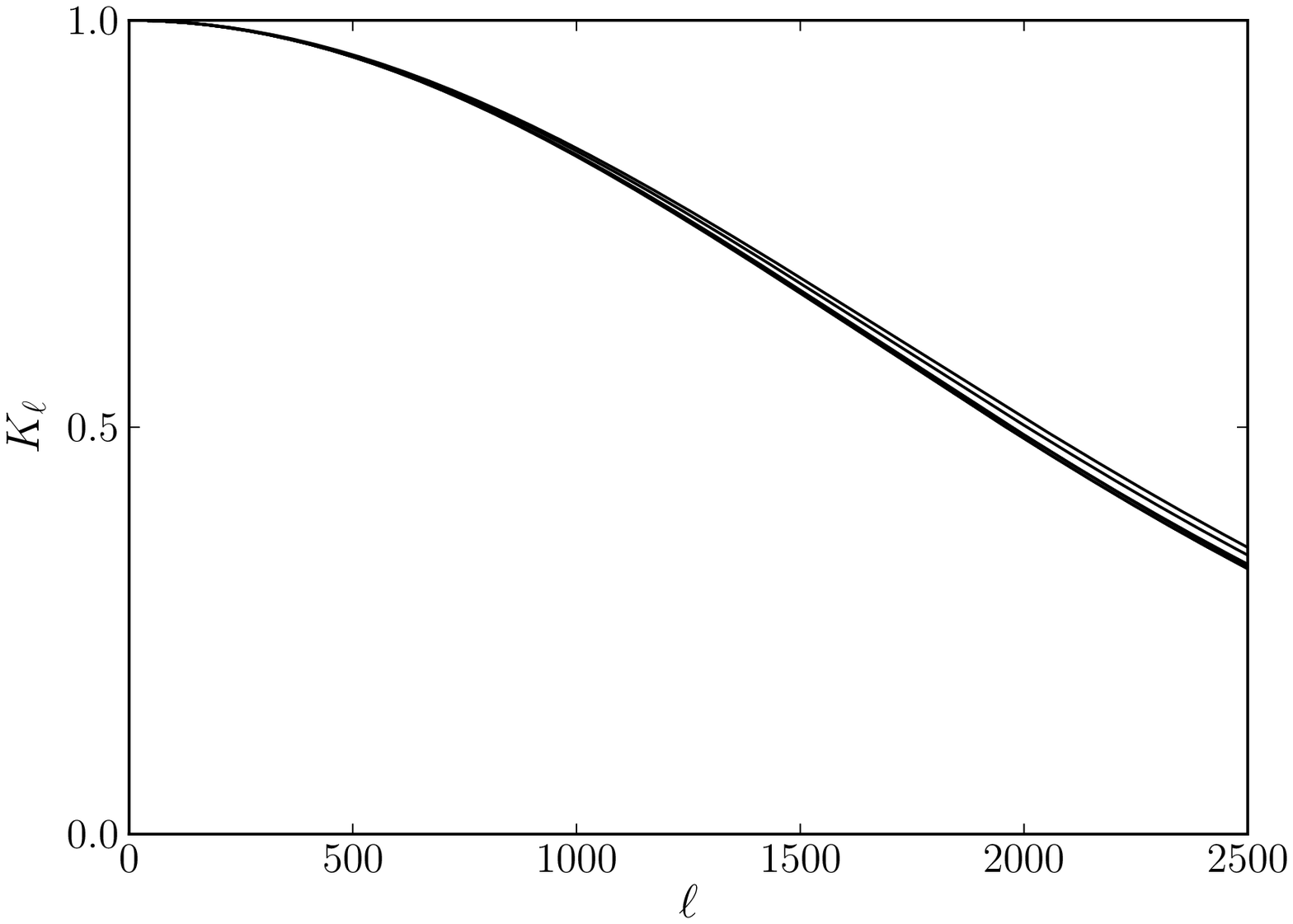}{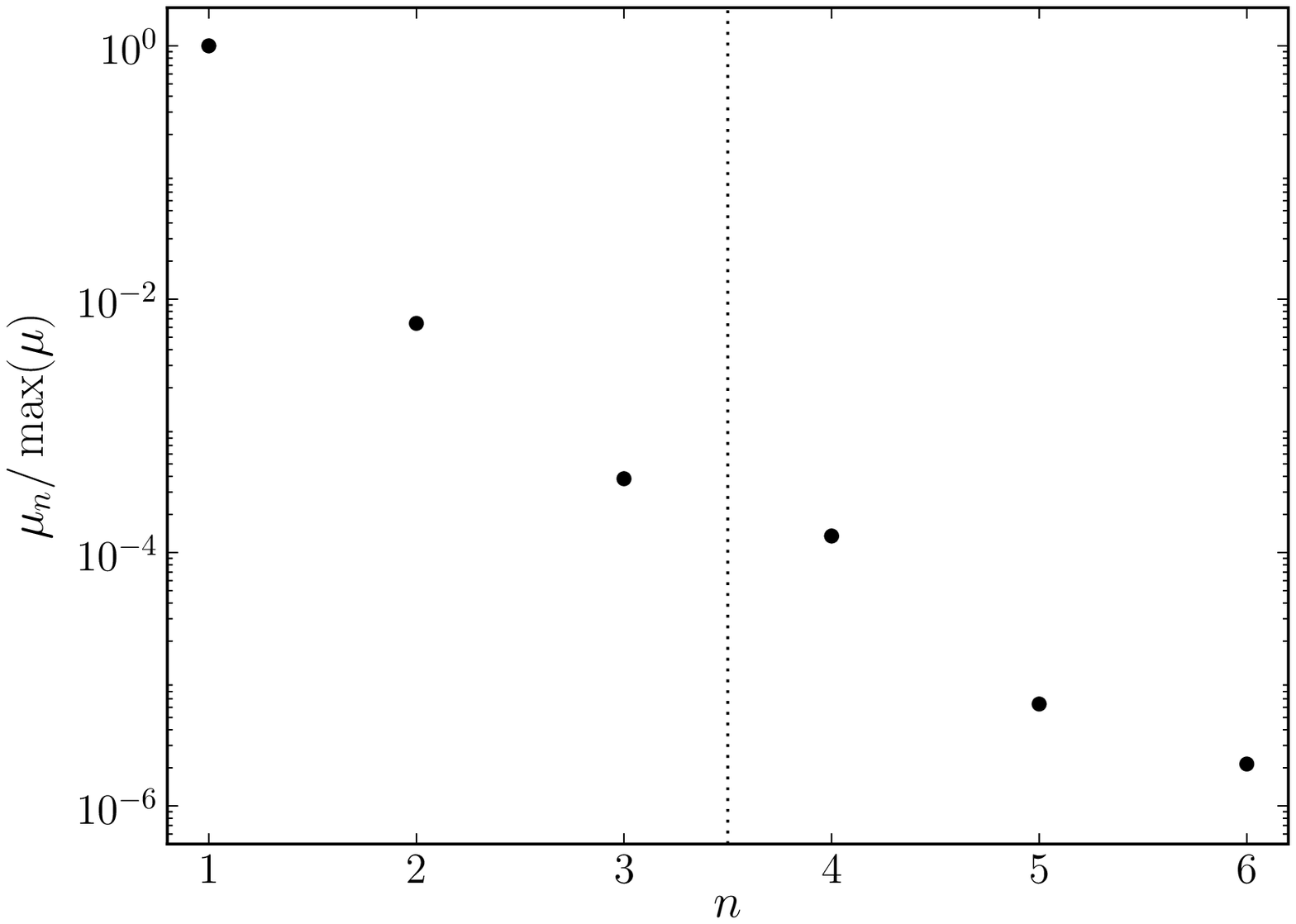}{fig:pl_beams}
{All six \planck\ beams at $217 \GHz$ (\emph{left panel}) have very
  similar shapes. As a result, the eigenvalues of their singular value
  decomposition decrease quickly (\emph{right panel}), allowing half
  of the modes to be safely discarded.}

\twofig{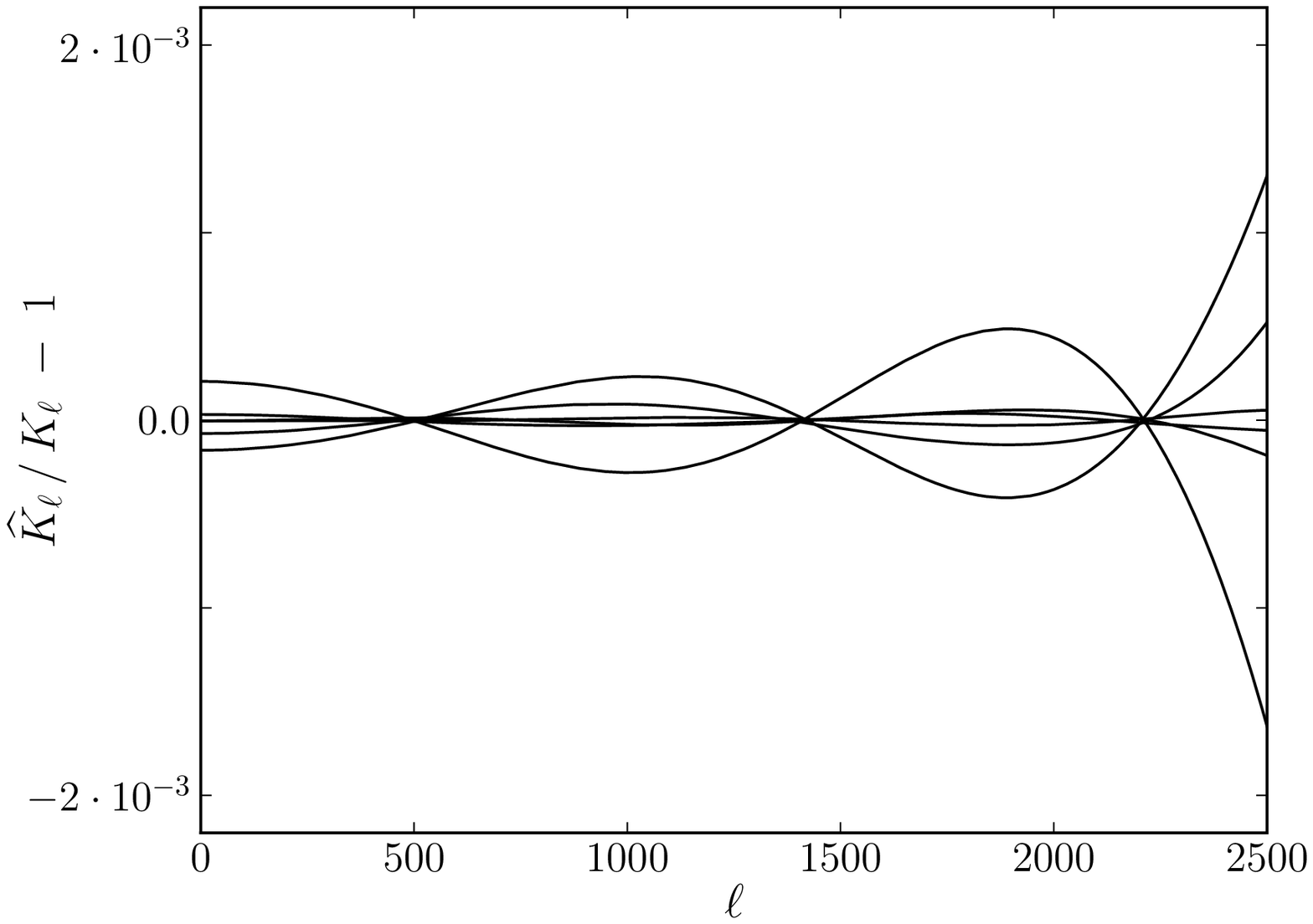}{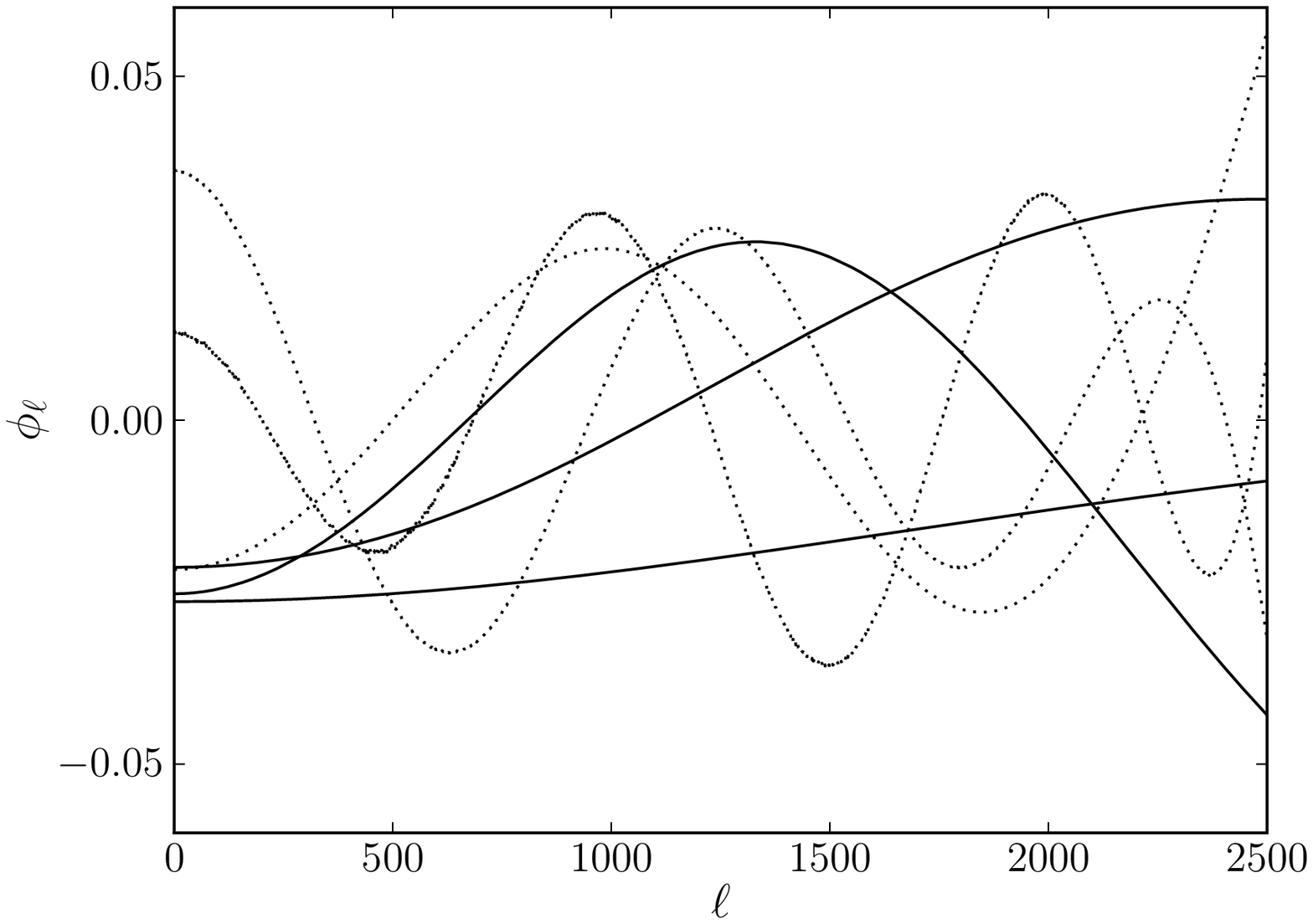}{fig:pl_modes1}
{\emph{Left panel:} Retaining the first three out of six \planck\ $217
  \GHz$ beam eigenmodes allows to reduce the relative truncation error
  of all convolution kernels to the order \order{10^{-3}}. \emph{Right
    panel:} We compare the eigenmodes used in the convolution
  (\emph{solid lines}) to the discarded modes (\emph{dotted
    lines}). Results in this plot have been obtained from a SVD,
  i.e.\ using kernel weights $(2 \ell + 1) \mathcal{C}_{\ell} /
  N^{(n)}_{\ell} = const.$}

\twofig{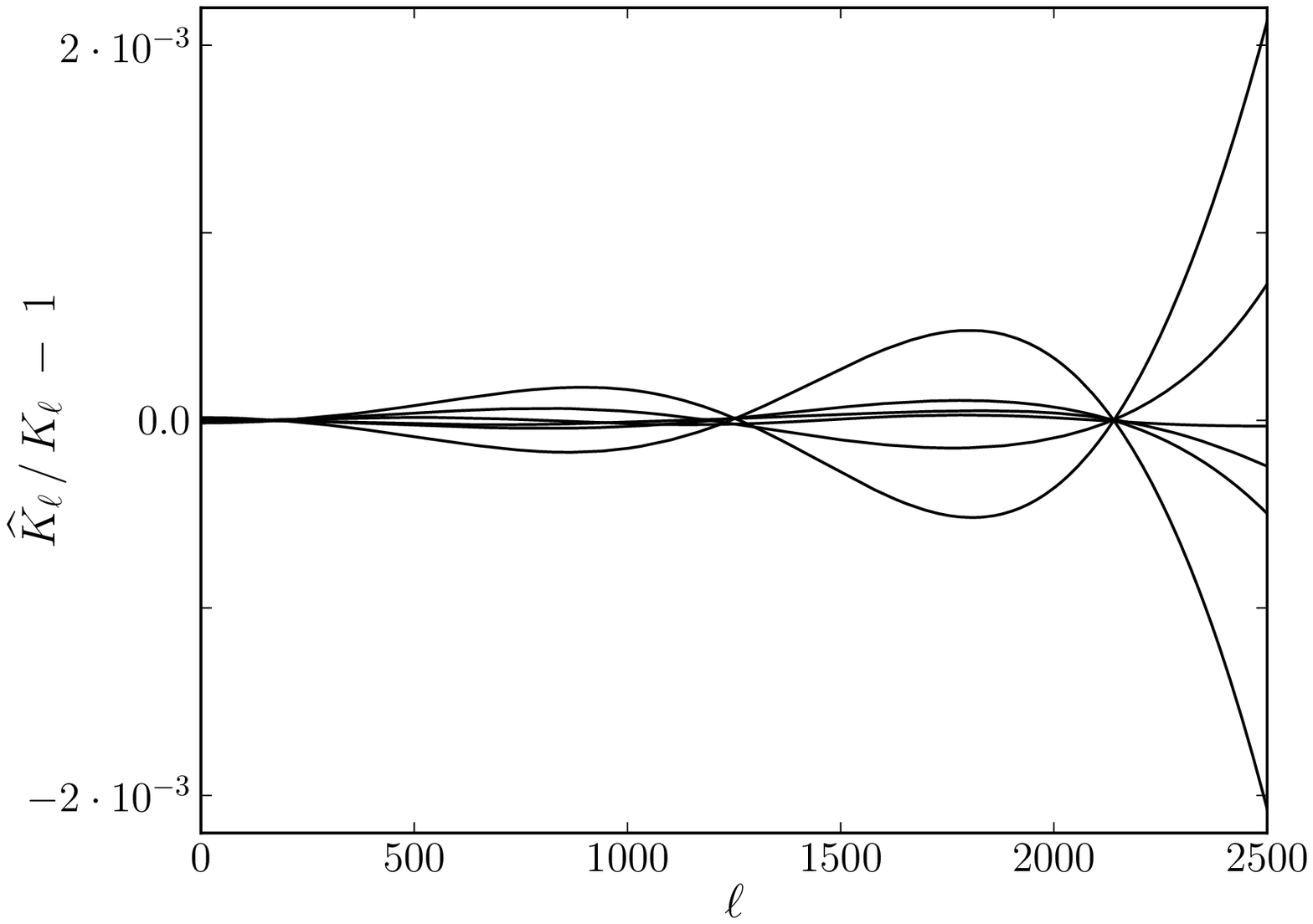}{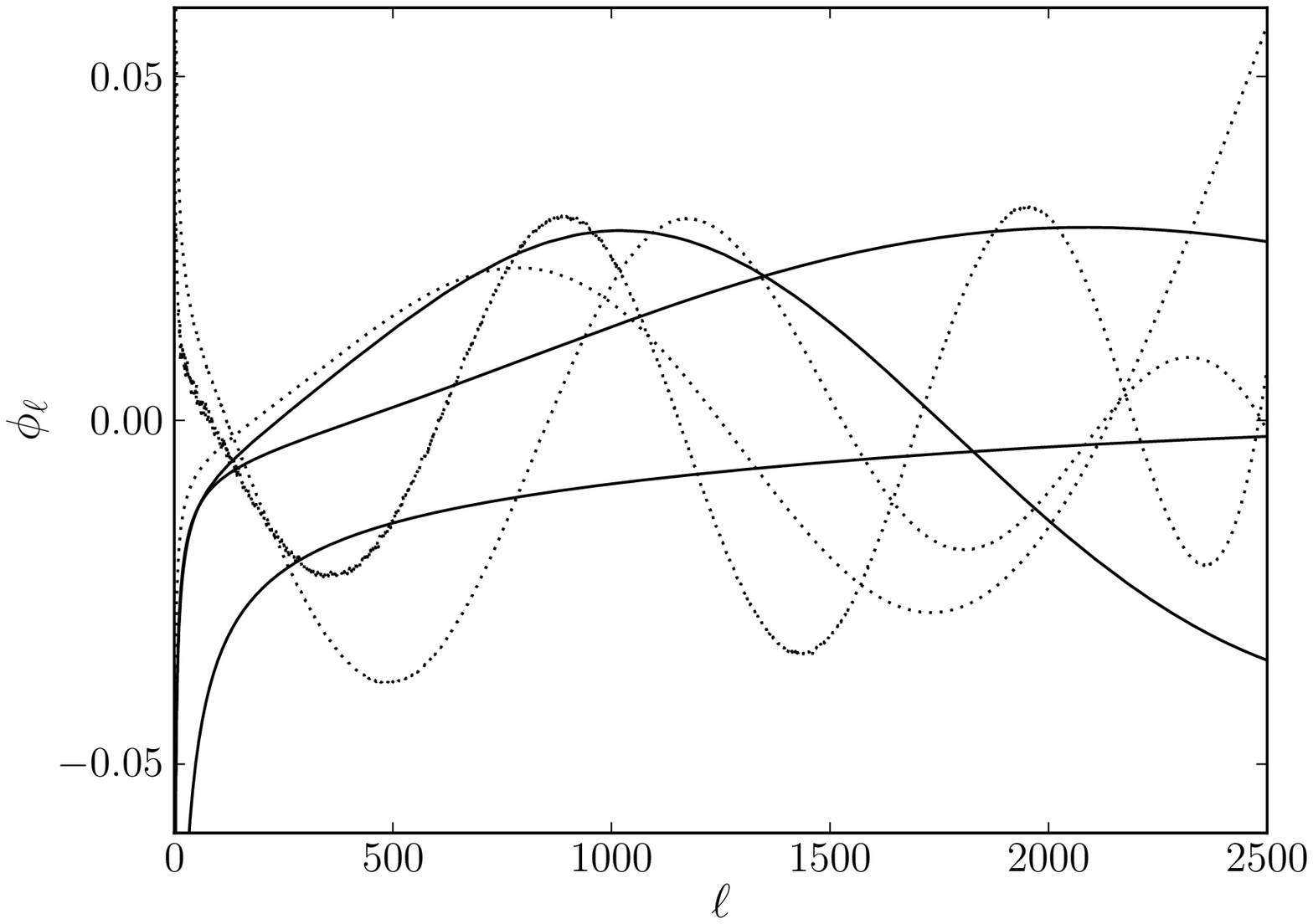}{fig:pl_modes2}
{Kernel weights allow for a full control over truncation errors. Same
  as \fig{fig:pl_modes1}, but for a $(2 \ell + 1) \mathcal{C}_{\ell} /
  N^{(n)}_{\ell} \propto (2 \ell + 1) / (\ell \, (\ell + 1))$
  weighting scheme, enforcing a more precise kernel reconstruction on
  large angular scales at the cost of increased errors at high
  multipoles.}

\threefig{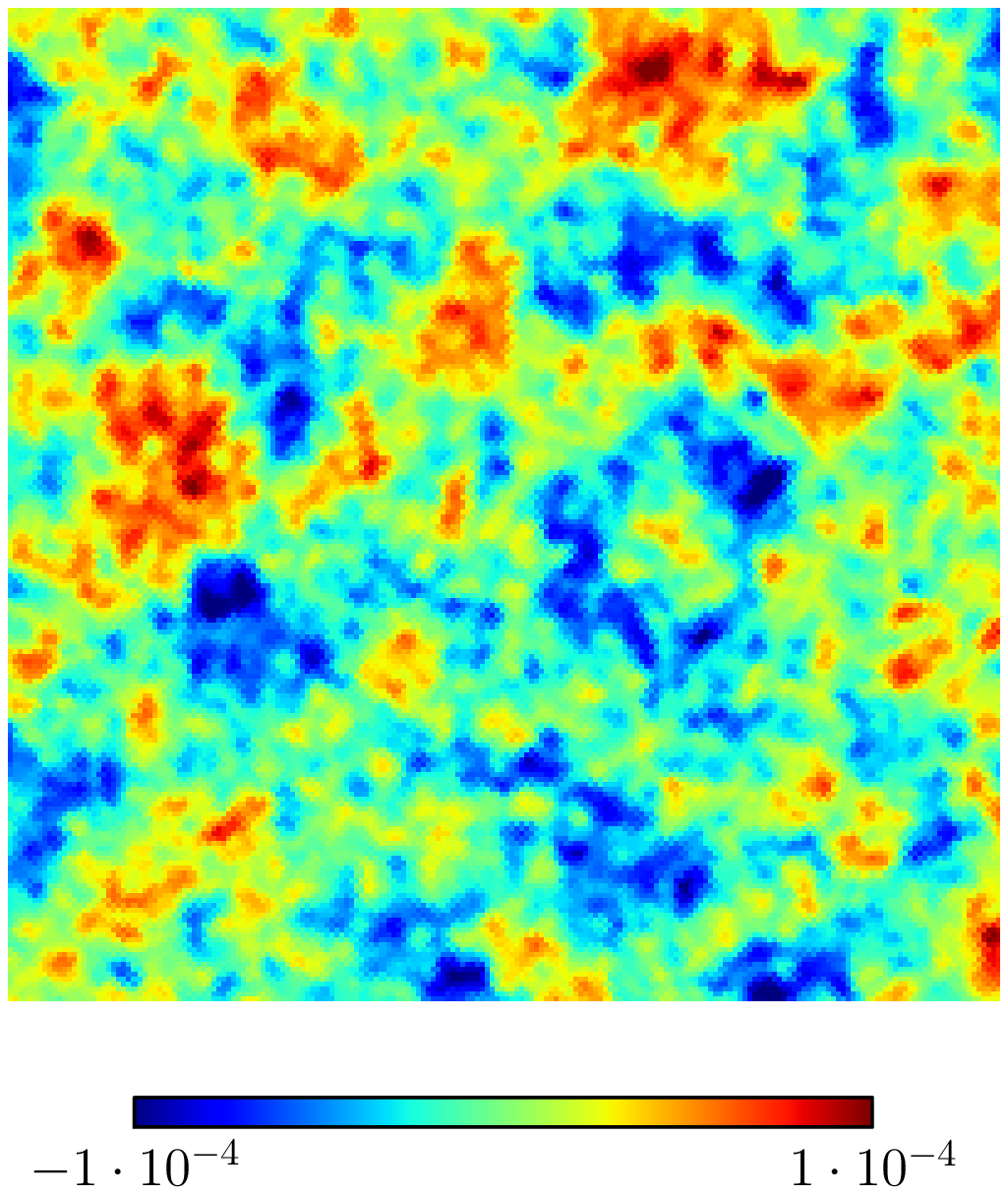}{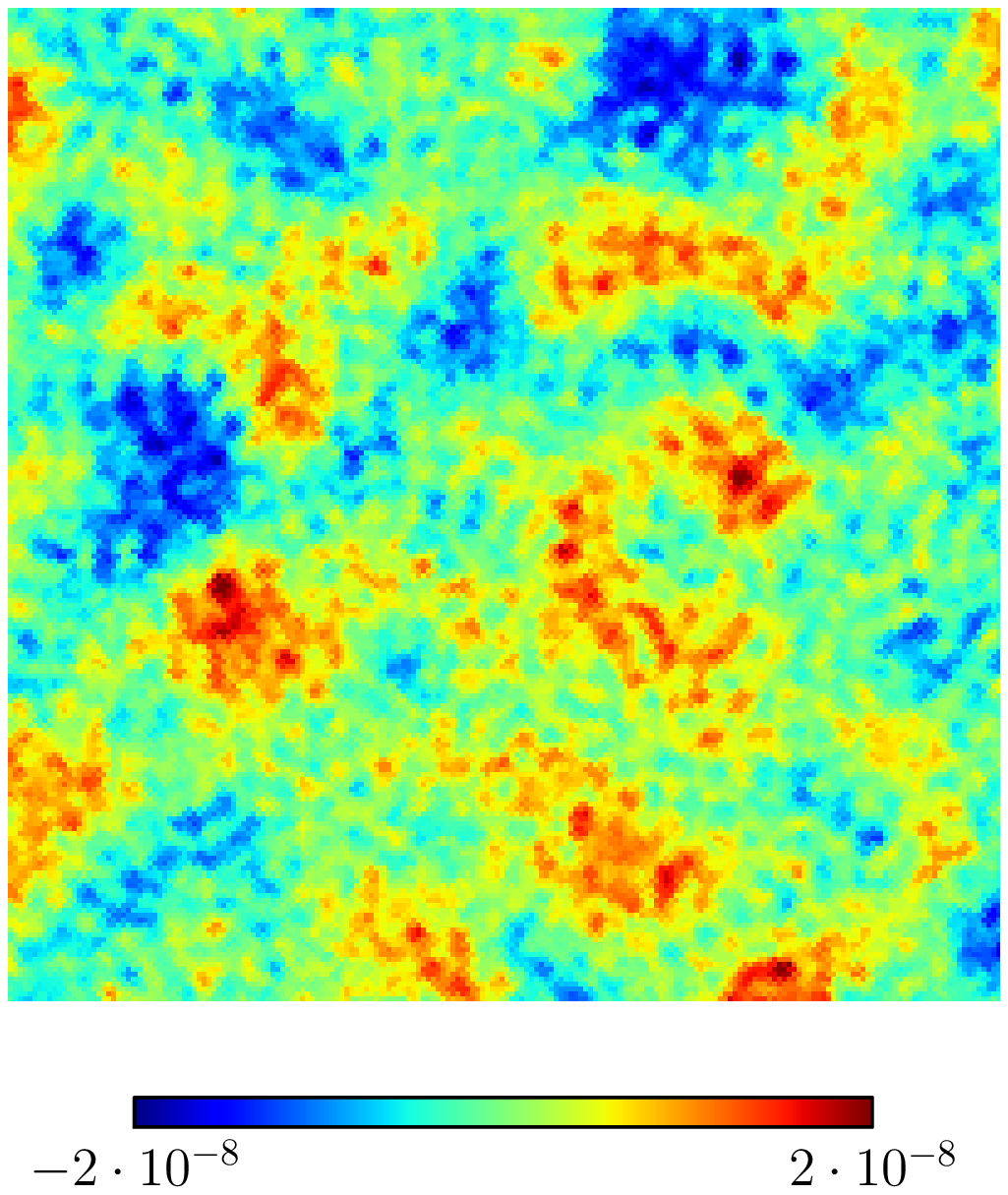}{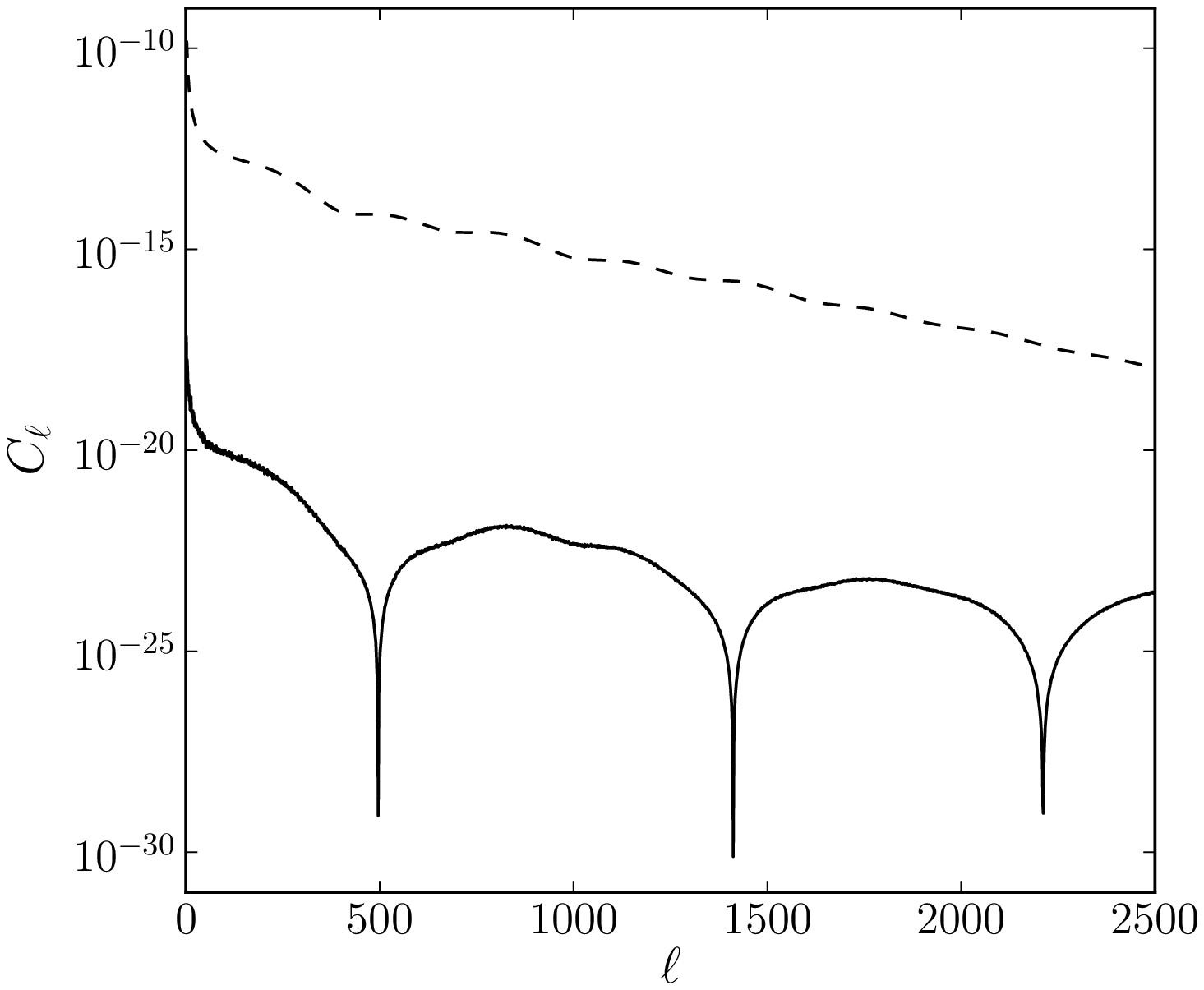}{fig:pl_error}
{Truncation errors are negligible. Using the kernel with the largest
  truncation error as worst case scenario, we plot the beam convolved
  simulated CMB map used in this test of the \planck\ $217 \GHz$
  channels (\emph{left panel}, we show a $10^{\circ} \times 10^{\circ}$
  patch). \emph{Middle panel:} The difference map between the results
  obtained with the exact convolution and the \cc\ with three beam
  modes. \emph{Right panel:} Compared to the fiducial power spectrum
  of the input map (\emph{dashed line}), the power spectrum of the
  difference map is subdominant by a large margin on all angular
  scales.}

The test demonstrates that the algorithm can be applied
straightforwardly to the beam convolution problem. In case of the six
\planck\ $217 \GHz$ detectors, we reduce the number of computationally
expensive spherical harmonic transformations by a factor of two. This
finding is characteristic for the scope of the algorithm: for a small
total number of convolution kernels, the reductions in computational
costs can only be modest. However, already for the latest generation
of CMB instruments, the \cc\ scheme can offer very large performance
improvements as we will demonstrate explicitly in the next paragraph.

\subsection{Keck}

Exemplary for modern ground based and balloon-borne CMB experiments, we
now discuss the application of the algorithm for the \keck\ array, a
polarization sensitive experiment located at the south pole that
started data taking in 2010 \citep{2010SPIE.7741E..50S}. Its
instrument currently consists of five separate receivers, each housing
496 detectors, and scanning the sky at a common frequency of $150
\GHz$.

Measurements have shown that the 2480 \keck\ beams can be described by
elliptic Gaussian profiles to good approximation
\citep{2012SPIE.8452E..26V},
\beq
\label{eq:k_beams}
K(\vec{x}) \propto \mathrm{e}^{-\frac{1}{2}(\vec{x} - \vec{x_0})
  \mathbf{C}^{-1} (\vec{x} - \vec{x_0})} \, ,
\eeq
where the beam center is located at $\vec{x_0}$. Here, the beam size
and ellipticity is parametrized by the covariance matrix,
\beq
\mathbf{C} = \sigma^2 \left( \begin{array}{cc}
  1 + \epsilon & 0 \\ 0 & 1 - \epsilon \end{array} \right) \, ,
\eeq
with the receiver specific parameters $\sigma$ and $\epsilon$
reproduced in \tab{tab:keck}.

To simulate the optical system of the full \keck\ array, we drew 2480
realizations of beam size and ellipticity according to the receiver
specifications and then used \eq{eq:k_beams} to construct individual
beams. We finally rotated the beams around their axes with randomly
chosen angles between $0 \le \phi < 2 \pi$. Applying fully random
rotations is conservative since beams of bolometers in the same
receiver are known to have similar orientations.

We found that only the first eight common eigenmodes are necessary to
approximate all 2480 individual beams to a precision of at least the
order \order{10^{-3}}. We illustrate this set of eigenmodes in
\fig{fig:k_eigenmodes}. In \fig{fig:k_reconstruction}, we show as an
example the beam with the largest reconstruction error. For about $90
\, \%$ of the detectors, the truncation errors are below
\order{10^{-4}}.

We verified the results with a CMB simulation in flat sky
approximation, high-pass filtered to suppress signal below $\ell <
50$. We plot the difference map computed from a direct convolution and
the \cc\ with eight eigenmodes in \fig{fig:k_error}. The error is
subdominant on all angular scales.

The example outlined here demonstrates the full strength of the
algorithm. Computing beam convolutions for the \keck\ array, we are
able to reduce the number of computationally expensive convolution
operations from 2480 to only eight, an improvement by a factor as high
as 310.

\begin{table}
  \centering
  \caption{\keck\ beam parameters as provided by
    \citet{2012SPIE.8452E..26V}.}
  \label{tab:keck}
  \begin{tabular}{l c c}
    \hline
    \hline
    & $\sigma / [^{\circ}]$ & $\epsilon$ \\
    \hline
    Receiver 0 & $0.214 \pm 0.005$ & $0.010 \pm 0.007$ \\
    Receiver 1 & $0.213 \pm 0.006$ & $0.012 \pm 0.006$ \\
    Receiver 2 & $0.213 \pm 0.006$ & $0.012 \pm 0.007$ \\
    Receiver 3 & $0.216 \pm 0.008$ & $0.013 \pm 0.010$ \\
    Receiver 4 & $0.218 \pm 0.013$ & $0.013 \pm 0.010$ \\
    \hline
  \end{tabular}
\end{table}

\onelargefig{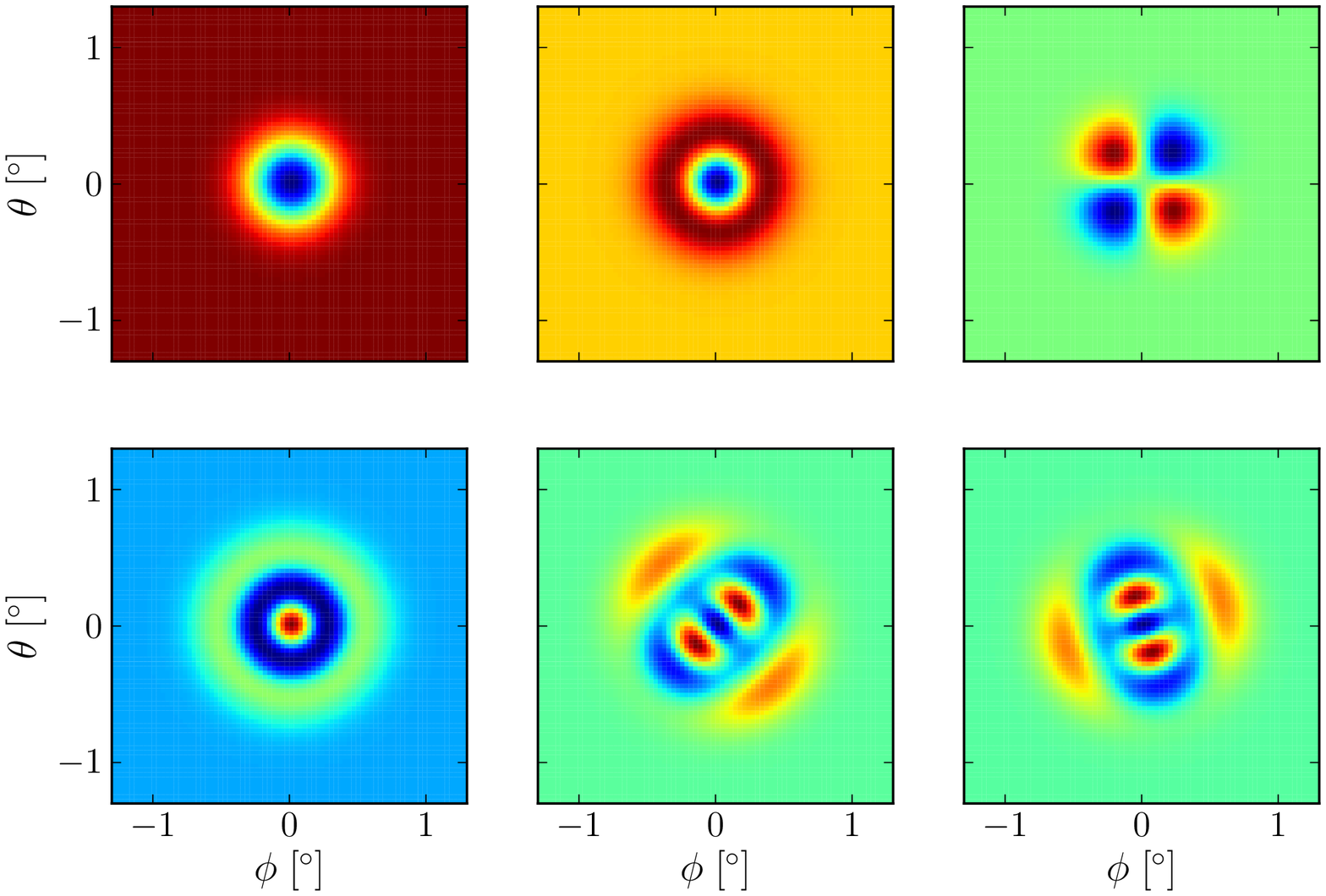}{fig:k_eigenmodes}
{Simulated 2480 asymmetric \keck\ beams at $150 \GHz$ are similar
  enough to be represented by only eight distinct beam eigenmodes to
  high precision.}

\twofig{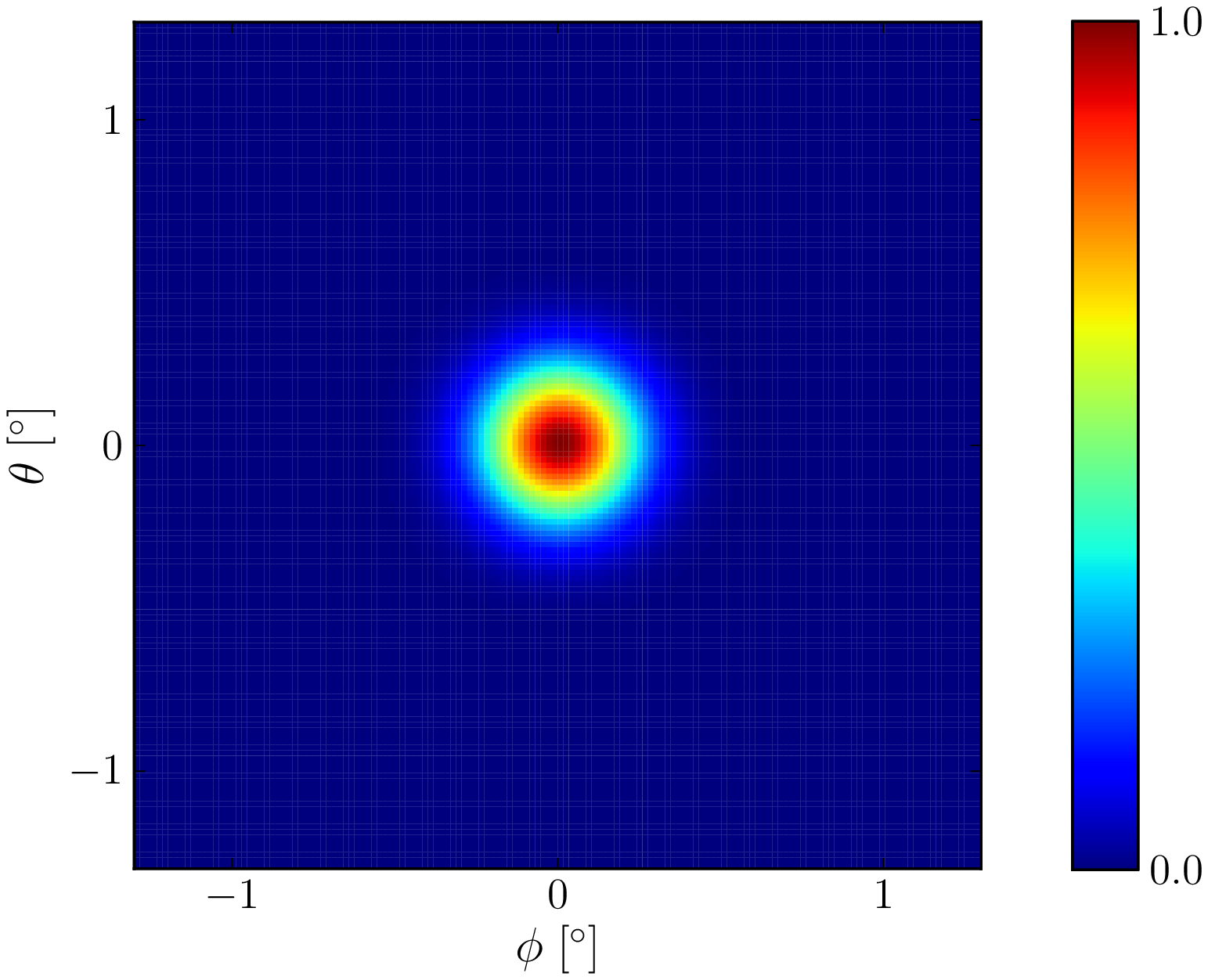}{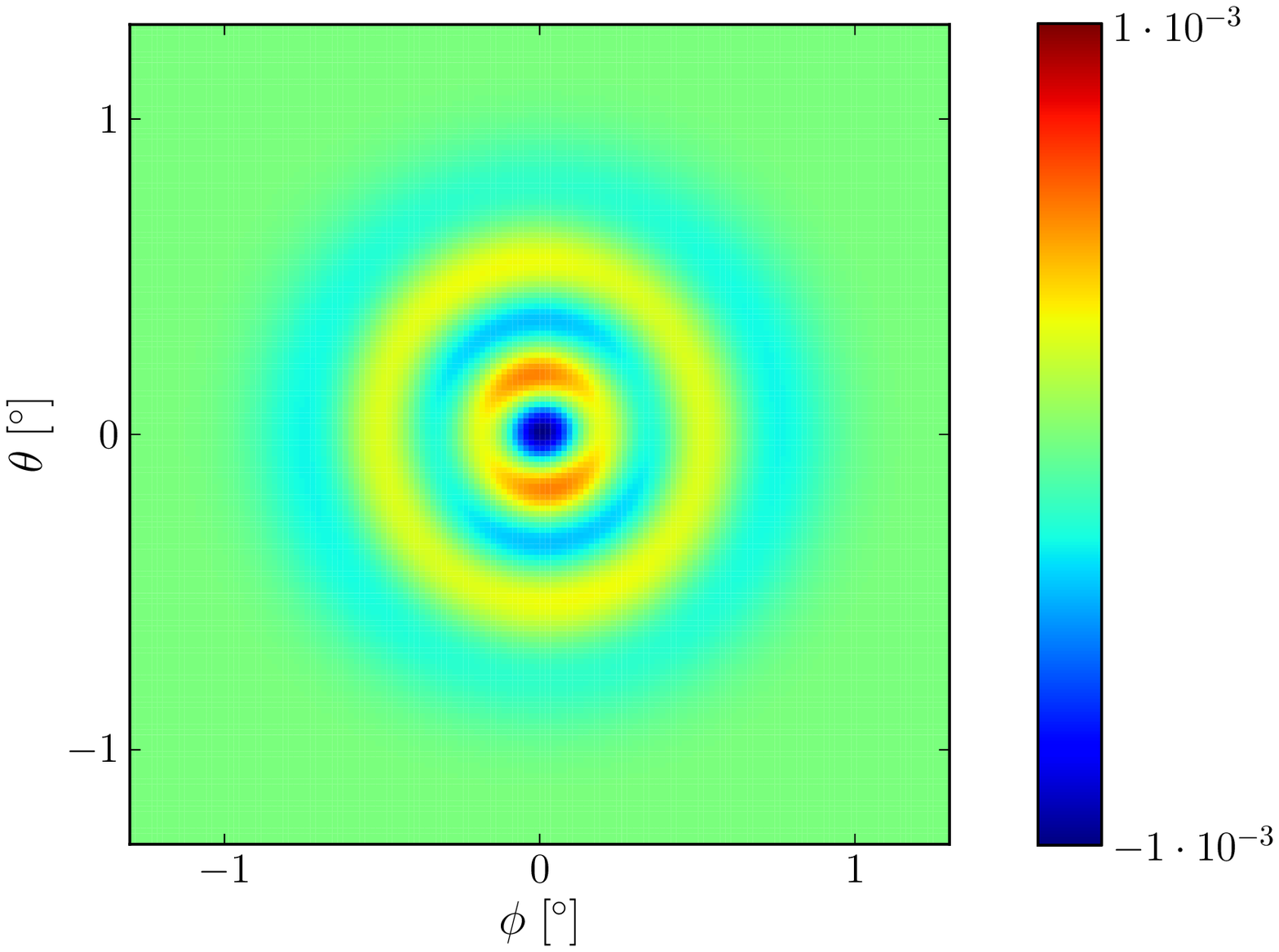}{fig:k_reconstruction}
{\emph{Left panel:} We show the beam with the largest reconstruction
  error for the simulated \keck\ array. \emph{Right panel:} Using the
  first eight beam eigenmodes, the truncation error is at most of the
  order \order{10^{-3}}.}

\threefig{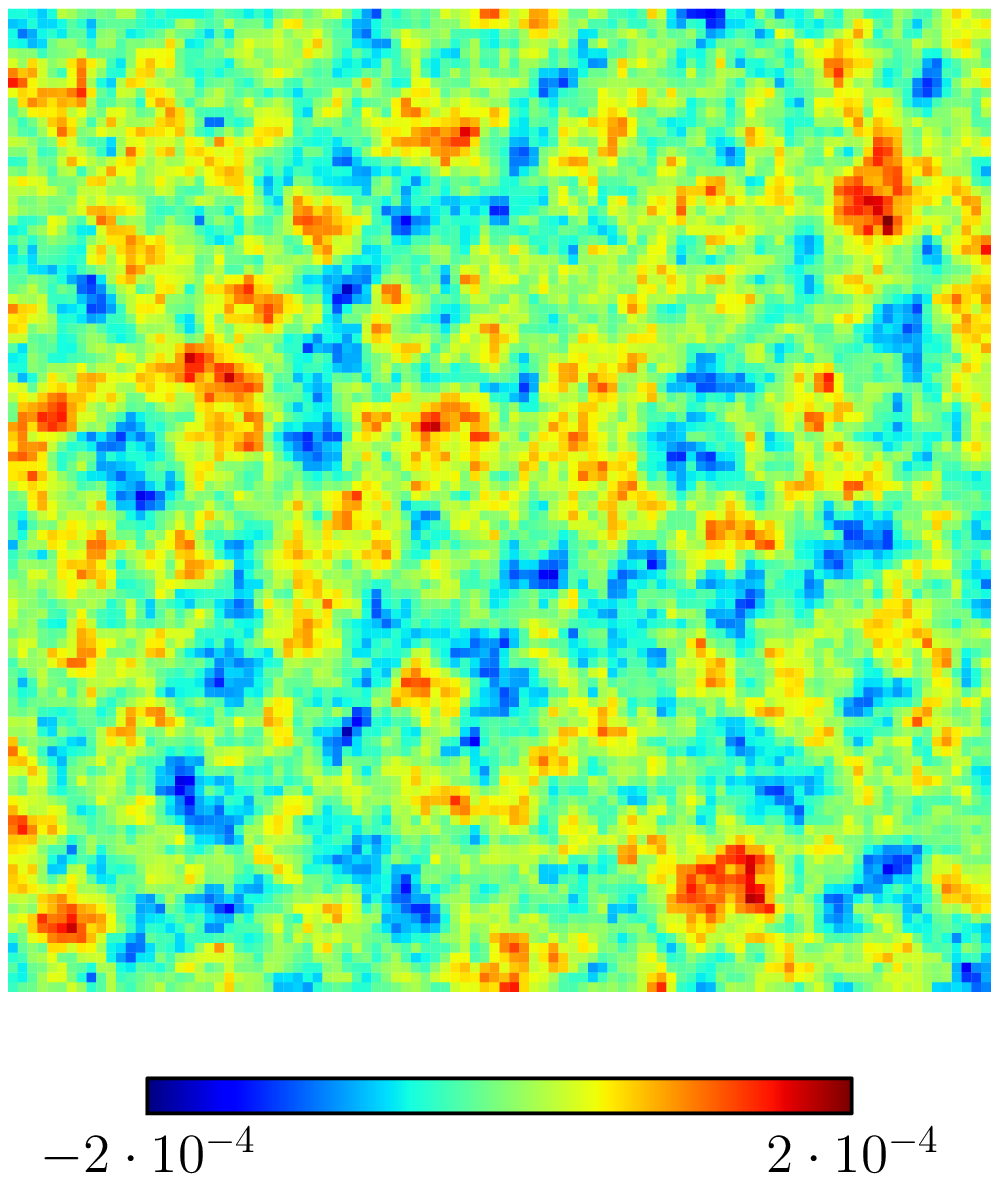}{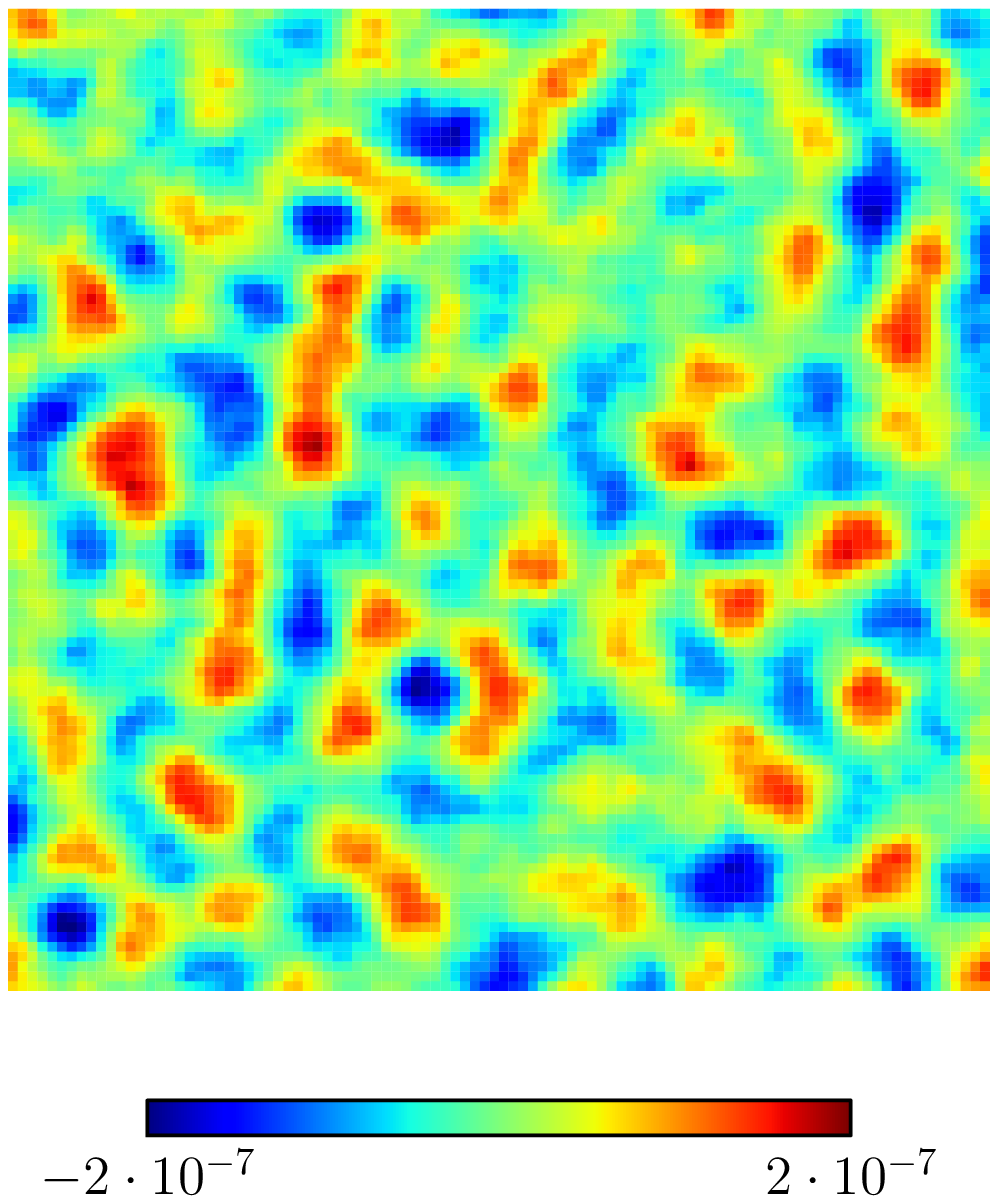}{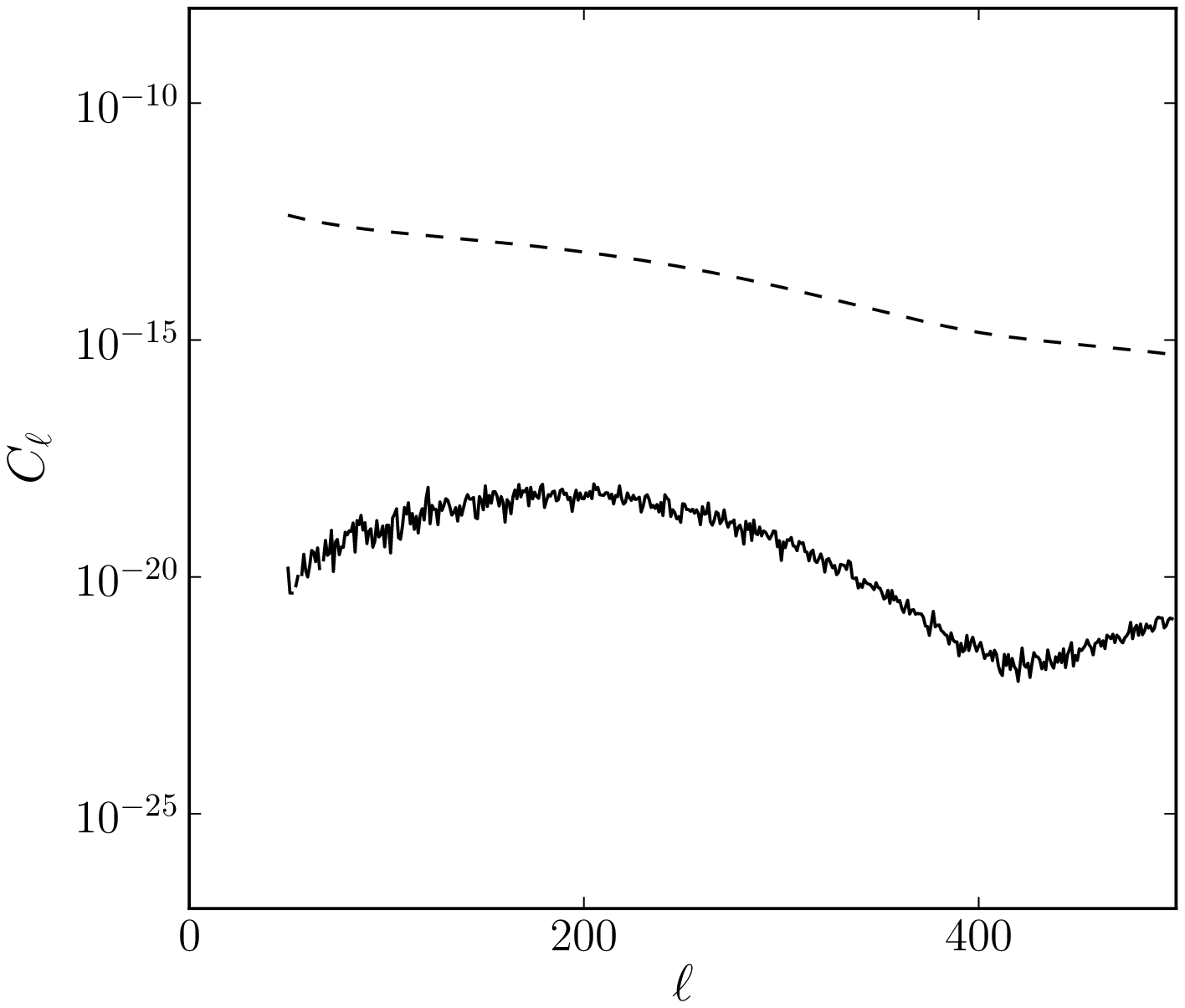}{fig:k_error}
{Same as \fig{fig:pl_error}, but for the worst case of the simulated
  \keck\ experiment. Using eight beam modes for the convolution is
  sufficient to reduce the truncation error to negligible levels.}

\section{Scope of the algorithm}
\label{sec:scope}

As shown in \sect{sec:example}, the algorithm has the potential to
provide huge speedups for the beam convolution operation of modern
experiments with a large number of detectors, necessary to improve the
sensitivity of CMB measurements in the photon noise limited
regime. Fast beam convolutions are not only important for the
simulation of signal maps for individual detectors. They also play a
crucial role in the mapmaking process, the iterative construction of a
common sky map out of the time ordered data from different detectors
observing at the same frequency.

Current experiments already deploy several hundreds to thousands of
detectors, making them ideal candidates for the algorithm, e.g.,
SPTpol (about 800 pixels, \citealt{2012SPIE.8452E..1EA}),
POLARBEAR (about 1300 pixels, \citealt{2012SPIE.8452E..1CK}),
EBEX (about 1400 pixels, \citealt{2010SPIE.7741E..37R}),
Spider (about 2600 pixels, \citealt{2010SPIE.7741E..46F}),
ACTPol (about 3000 pixels, \citealt{2010SPIE.7741E..51N}). For
future experiments, the number of detectors can be expected to
increase further, e.g., for PIPER (about 5000 pixels,
\citealt{2013AAS...22122904L}), the Cosmic Origins Explorer
(about 6000 pixels, \citealt{2011arXiv1102.2181T}), or
POLARBEAR-2 (about 7500 pixels \citealt{2012SPIE.8452E..1HT}),
making the application of the algorithm even more rewarding.

The new method also allows a fast implementation of matched filtering
on the sphere (or other domains) if the size of the target is unknown
\citep[e.g., to detect signatures of bubble collisions in the
  CMB,][]{2012PhRvD..85j3502M}, or analogously for continuous wavelet
transforms, frequently used in the context of data compression or
pattern recognition \citep[e.g.,][]{1989ieee...7M}. Here, the input
signal is convolved with a large set of scale dilations of an
analyzing filter or wavelet. Since the resulting convolution kernels
are of similar shape by construction, the decomposition into only a
few eigenmodes can be done efficiently. Our new method therefore has
the potential to increase the numerical performance of such
computations by a substantial factor.

Finally, besides from the reduction in computational costs, we note
that \cc\ may also offer the possibility to reduce the disk space
required to store convolved data sets. Instead of saving the convolved
signal for each kernel separately, it now becomes possible to just
keep the compressed output for the most important eigenmodes, and
efficiently decompress it with their proper weights for each
individual kernel on the fly as needed.

\section{Summary}
\label{sec:summary}

In signal processing, a single data set often has to be convolved with
many different kernels. With increasing data size, this operation
quickly becomes numerically expensive to evaluate, possibly even
dominating the execution time of analysis pipelines.

To increase the performance of such convolution operations, we
introduced the general method of \cc. Using an eigenvector
decomposition of the convolution kernels, we first obtain their
optimal expansion into a common set of basis functions. After ordering
the modes according to their relative importance, we identify the
minimal number of basis functions to retain to satisfy the accuracy
requirements. Then, the convolution operation is executed for each
mode separately, and the final result obtained for each kernel from a
linear combination.

This algorithm offers particularly large performance improvements, if
\begin{itemize}
\item the total number of kernels to consider is large, and
\item the kernels are sufficiently similar in shape, such that they
  can be approximated to good precision with only a few eigenmodes.
\end{itemize}

In case of the analysis of CMB data, we use the beam convolution
problem as an example application of the \cc\ scheme. On the basis of
simulations of the \keck\ array with 2480 detectors
\citep{2012SPIE.8452E..26V}, we demonstrated that the \cc\ scheme
allows to reduce the number of beam convolution operations by a factor
of about 300, offering the possibility to cut the runtime of
convolution pipelines by orders of magnitude. Additional improvements
are possible when used in combination with efficient convolution
algorithms \citep[e.g.,][]{2011A&A...532A..35E}.

\begin{acknowledgements}
The authors thank Clem Pryke for highlighting beam convolution for
kilo-detector experiments as an outstanding problem, and Guillaume
Faye and Xavier Siemens for conversations regarding the applications
to searches for gravitational wave signals. BDW was supported by the
ANR Chaire d'Excellence and NSF grants AST 07-08849 and AST 09-08902
during this work. FE gratefully acknowledges funding by the CNRS. Some
of the results in this paper have been derived using the HEALPix
\citep{2005ApJ...622..759G} package.

Based on observations obtained with \planck\
(\url{http://www.esa.int/Planck}), an ESA science mission with
instruments and contributions directly funded by ESA Member States,
NASA, and Canada.

The development of \planck\ has been supported by: ESA; CNES and
CNRS/INSU-IN2P3-INP (France); ASI, CNR, and INAF (Italy); NASA and DoE
(USA); STFC and UKSA (UK); CSIC, MICINN and JA (Spain); Tekes, AoF and
CSC (Finland); DLR and MPG (Germany); CSA (Canada); DTU Space
(Denmark); SER/SSO (Switzerland); RCN (Norway); SFI (Ireland);
FCT/MCTES (Portugal); and The development of \planck\ has been supported
by: ESA; CNES and CNRS/INSU-IN2P3-INP (France); ASI, CNR, and INAF
(Italy); NASA and DoE (USA); STFC and UKSA (UK); CSIC, MICINN and JA
(Spain); Tekes, AoF and CSC (Finland); DLR and MPG (Germany); CSA
(Canada); DTU Space (Denmark); SER/SSO (Switzerland); RCN (Norway);
SFI (Ireland); FCT/MCTES (Portugal); and PRACE (EU).

A description of the \planck\ Collaboration and a list of its members,
including the technical or scientific activities in which they have
been involved, can be found at
\url{http://www.sciops.esa.int/index.php?project=planck&page=Planck_Collaboration}.

\end{acknowledgements}

\bibliographystyle{aa}
\bibliography{literature}

\end{document}